\documentclass[11pt]{article}
	
	\newcommand{\blind}{0}
	
    \makeatletter
    \renewcommand\section{\@startsection {section}{1}{\z@}%
                                       {-3.5ex \@plus -1ex \@minus -.2ex}%
                                       {2.3ex \@plus.2ex}%
                                       {\normalfont\fontfamily{phv}\fontsize{16}{19}\bfseries}}
    \renewcommand\subsection{\@startsection{subsection}{2}{\z@}%
                                         {-3.25ex\@plus -1ex \@minus -.2ex}%
                                         {1.5ex \@plus .2ex}%
                                         {\normalfont\fontfamily{phv}\fontsize{14}{17}\bfseries}}
    \renewcommand\subsubsection{\@startsection{subsubsection}{3}{\z@}%
                                        {-3.25ex\@plus -1ex \@minus -.2ex}%
                                         {1.5ex \@plus .2ex}%
                                         {\normalfont\normalsize\fontfamily{phv}\fontsize{14}{17}\selectfont}}
    \makeatother
	\usepackage{amsmath}
	\usepackage{graphicx}
	\usepackage{enumerate}
	\usepackage{xcolor}
    \usepackage{booktabs}
    \usepackage{threeparttable}
    \usepackage{longtable}
    \usepackage[utf8]{inputenc}
    \usepackage[style=apa]{biblatex}
    \usepackage{xr} % refereces to other documents (SI)
    \usepackage{url} % not crucial - just used below for the URL
    \usepackage{lineno}
    \usepackage[final]{changes} % for final version

\begin{document}
			%%%%%%%%%%%%%%%%%%%%%%%%%%%%%%%%%%%%%%%%%%%%%%%%%%%%%%%%%%%%%%%%%%%%%%%%%%%%%%
		\def\spacingset#1{\renewcommand{\baselinestretch}%
			{#1}\small\normalsize} \spacingset{1}
		%%%%%%%%%%%%%%%%%%%%%%%%%%%%%%%%%%%%%%%%%%%%%%%%%%%%%%%%%%%%%%%%%%%%%%%%%%%%%%
		
		\if0\blind
		{
			\title{\bf Website visits can predict angler presence using machine learning}
			\author{Julia S. Schmid\thanks{Corresponding author (email: jul.s.schmid@gmail.com)} $^1$, Sean Simmons$^2$, Mark A. Lewis$^{1,3,4,5}$, Mark S. Poesch$^6$, Pouria Ramazi$^7$ \\
			\small
			   $^1$Department of Mathematical and Statistical Sciences, University of Alberta, Edmonton, Alberta, Canada \\
			 \small
             $^2$Angler’s Atlas, Goldstream Publishing, Prince George, British Columbia, Canada \\
             \small
             $^3$Department of Mathematics and Statistics, University of Victoria, Victoria, British Columbia, Canada \\
             \small
             $^4$Department of Biology, University of Victoria, Victoria, British Columbia, Canada \\
             \small
             $^5$Department of Biological Sciences, University of Alberta, Edmonton, Alberta, Canada \\
             \small
             $^6$Department of Renewable Resources, University of Alberta, Edmonton, Alberta, Canada \\
             \small 
             $^7$Department of Mathematics and Statistics, Brock University, St. Catharines, Ontario, Canada}
			\date{}
			\maketitle
		} \fi
		
		\if1\blind
		{

            \title{\bf \emph{First Manuscript Anglers Project}}
			\author{Author information is purposely removed for double-blind review}
			
\bigskip
			\bigskip
			\bigskip
			\begin{center}
				{\LARGE\bf \emph{First Manuscript Anglers Project}}
			\end{center}
			\medskip
		} \fi
		\bigskip

%    \linenumbers
    \newpage	
    \spacingset{2}
	\begin{abstract}

%\textcolor[rgb]{0.00,0.07,1.00}{We strongly encourage authors to address the following three questions in their \textbf{abstract}, preferably following the order shown: (1) Research problem statement: what is the research problem to be addressed? (2) Methods and results: how do the authors address the research problem and what are the main results? (3) Insights and implications: What have the authors learned (as opposed to what they did, which is covered in point (2)) from conducting this research? What is the knowledge gained and why does it matter?\\

Understanding and predicting recreational \replaced{angler effort}{fishing activity} is important for sustainable fisheries management. 
However, \replaced{conventional}{traditional} methods of measuring \replaced{angler effort}{fishing pressure}, such as surveys, can be costly and limited in both time and spatial extent. 
\replaced{M}{Predictive m}odels that \replaced{predict angler effort}{relate fishing activity} \replaced{based on}{to} environmental or economic factors typically rely on historical data, which often \replaced{limits}{restricts} their spatial \replaced{and temporal generizability}{applicability} due to data scarcity.
In this study, high-resolution \deleted{angler-generated }data from an online \added{fishing} platform and easily accessible auxiliary data were tested to predict daily boat presence and aerial counts of boats at almost 200 lakes over five years in Ontario, Canada. 
Lake-information website visits alone enabled predicting daily angler boat presence with 78\% accuracy. 
While incorporating additional environmental, socio-ecological, weather and angler-\replaced{reported}{generated} features into machine learning models did not remarkably improve prediction performance of boat presence, they were substantial for the prediction of boat counts. 
%Daily boat counts were predicted with an R² of 0.78, and the presence or absence of fishing boats was predicted with XX\% accuracy. 
Models achieved an R² of up to 0.77 at known lakes included in the model training, but they performed poorly for unknown lakes (R² = 0.21).
The results demonstrate the value of integrating \deleted{angler-generated }data from online \added{fishing} platforms into predictive models and highlight the potential of machine learning models to enhance fisheries management.

%Heavy (Increasing?) angler pressure is leading to overfishing and outbreaks of dangerous fish diseases in some Canadian water bodies. Predicting angler's behavior in time and space is an essential task for the conservation of fish populations by, e.g., introducing new regulations on time. However, collecting required data by surveys is very costly. Recreational anglers can use websites and apps to report their trips. This citizen data provides a promising basis for predicting angler's behaviour. (However, it is biased to a specific group of anglers (Jiorie 2016).) In this study, we apply Bayesian networks to extrapolate reported angler data in space using environmental and demographic covariates. The approach enabled us to predict 5\% of the total angler behavior in three provinces of Canada. Anglers reporting their trips in the app prefer to go to ... [Key result of the map that we will show in the results]. Our study shows the potential of Bayesian networks combined with citizen data to predict the behavior of all anglers worldwide.
	\end{abstract}
			
	\noindent%
	{\it Keywords:} Angler-\replaced{reported}{generated} data; \replaced{angler effort}{angler pressure}, recreational fishing, freshwater fishing, boat counts, spatio-temporal prediction.

	%\newpage
	\spacingset{2} % DON'T change the spacing!

\newpage

\section{Introduction} \label{s:intro}
Recreational fisheries play a central role in the environmental, economic and social context of many regions (\cite{fao2020recreational, arlinghaus2017understanding}). 
Data on angler effort \replaced{can provide valuable insights}{are essential }for \deleted{effective }fisheries management, conservation strategies, and \added{the} sustainable use of aquatic resources\added{, particularly when used to estimate harvest rates and understand pressure on fish populations (\cite{brownscombe2019future, collins2022dynamic, slaton2023explaining})}. 
Understanding angler behavior in time and space enables \replaced{broad-scale}{regional-level} management, helping to allocate resources efficiently and mitigate potential negative impacts on fish populations and ecosystems (\cite{matsumura2019ecological, askey2013linking, cooke2005we, arlinghaus2017understanding}). 
Furthermore, predicting future angler behavior can aid in preparing for changes driven by environmental, socio-economic, and climatic factors (\cite{OMNRF2023, maldonado2024water, rijnsdorp2009resolving}).

Conventional methods, including on-site surveys and aerial counts, can be used to measure angler effort \added{(\cite{pollock1997catch, morrow2022improving})}. 
In Canada, fish stocks and the behavior of anglers are monitored by various institutions at regular time intervals. 
For example, Fisheries and Oceans Canada (DFO) runs a Canada-wide mail survey every five years to collect information on activities related to recreational fishing (\cite{DFO2019}). 
Similarly, the Ministry of Natural Resources and Forestry in Ontario conducts recreational fishing mail surveys every five years, and an annual fish community index gill netting program at Lake Ontario and Bay of Quinte (\cite{hunt2022selected, OMNRF2023}). 
%The Alberta Conservation Association's creel survey program involves summer-season surveys of anglers' catch and effort on four fisheries with over-harvest of walleye and northern pike (\cite{walker2007effectiveness}). 
The surveys are instrumental in assessing \replaced{angler effort}{fishing pressure}, understanding seasonal trends, and guiding management actions such as stocking and habitat restoration.
However, surveys are typically conducted at specific times and in specific locations, and are limited by logistical and financial considerations, so they may not capture the full variability of angler \replaced{effort}{activity} throughout the year or in different areas (\cite{smallwood2012expanding, wise2013determination, morrow2022improving, alexiades2015measurement}). 

To complement and extend the insights gained from conventional surveys, various models have been developed and tested to predict angler behavior and fill \deleted{the enormous} gaps in spatial and temporal coverage \added{(\cite{askey2013linking, jensen2022phenomenological, trudeau2021estimating})}.
Statistical models, such as simple regressions and generalized linear models, have used historical data to identify patterns and predict future angler effort (\cite{van2015imputing, trudeau2021estimating, askey2018angler, smith2024seasonal, beard2003impacts, mee2016interaction}). 
Dynamic models, including agent-based models and spatio-temporal models, provided enhanced predictive capabilities by simulating interactions between anglers and their environment over time (\cite{askey2013linking, post2008angler}). 
%These models have demonstrated the potential to predict and understand angler effort with varying degrees of accuracy, depending on the quality, granularity and type (e.g., stocking events, fish sizes) of the input data.
Commonly used factors in the predictive models include environmental variables (e.g., lake size, weather conditions, fish sizes), socio-ecological variables (e.g., population density, accessibility), management variables (e.g., harvest regulations, stocking events) and historical fishing data (\cite{hunt2019predicting, kane2020spatial, solomon2020frontiers, matsumura2019ecological, post2008angler, askey2013linking, beard2003impacts}). %While these factors provide a robust foundation for prediction, models often struggle with incorporating real-time and fine-scale data, leading to potential inaccuracies in dynamic and rapidly changing environments.

Despite advancements, current models\deleted{ and factors} used \replaced{to predict}{in predicting} angler behavior \replaced{face}{have notable } limitations \added{largely due to the availability and quality of input data}. 
\added{While many modeling approaches are capable of integrating real-time and high-resolution spatiotemporal data, such data are often unavailable, inconsistent, or not scientifically validated, which constrains model performance.}
\replaced{Moreover, a heaviy reliance}{Most exosting models lack the ability to integrate real-time data and often rely} on historical data\deleted{, which} may not accurately reflect current conditions, leading to potential inaccuracies in dynamic and rapidly changing environments.
Additionally, the spatial and temporal resolution of \replaced{many existing}{available} data \replaced{is limited}{can be insufficient.} \added{which can reduce the ability of models to capture fine-scale variation in angler behavior.}
\replaced{For example, s}{S}ome models focus only on temporal dynamics and disregard \replaced{spatial heterogeneity in}{varied} fishing effort \deleted{across space }(\cite{solomon2020frontiers, howarth2024non}).

A new and innovative way to collect more \added{timely and spatially detailed} data is the use of \deleted{citizen-reported} data from online platforms and fishing mobile applications as they provide a valuable and easily accessible source of information (\cite{venturelli2017angler, gundelund2022investigating, johnston2022comparative, gundelund2021changes}). 
These data can complement \replaced{conventional}{traditional} data sources by \replaced{offering higher resolution in both time and space}{providing timely, location-specific insights into angler activities, preferences, and environmental} \added{and have the potential to capture dynamic behavioral responses to rapidly changing environmental and social} conditions (\cite{johnston2022comparative, gundelund2022investigating}). 
\deleted{Moreover, machine learning (ML) methods such as random forests, gradient boosting, and neural networks can leverage these rich and diverse data sources to enhance predictions by identifying complex, non-linear relationships between environmental variables and angler behavior (\cite{breiman2001random, friedman2001greedy, goodfellow2016deep}).}
\added{Specifically, such data could allow near real-time monitoring of angler effort, detect deviations from expected angler behavior patterns, and reveal short-term trends that might otherwise be missed.}

\added{However, the high volume and heterogeneity of angler-reported data pose challenges for analysis.}
\added{In this context, machine learning (ML) methods such as random forests, gradient boosting, and neural networks are well-suited to leverage these rich and complex datasets.} 
\added{ML models can uncover non-linear and interactive effects among variables, identify patterns that may not be captured by conventional statistical approaches, and generate timely predictions (\cite{breiman2001random, friedman2001greedy, goodfellow2016deep}).}

In this study, easily accessible, real-time environmental, socio-ecological, and weather variables, along with angler-reported data from an online fishing platform, were used to train several ML models for predicting the spatio-temporal dynamics of angler effort measured by aerial surveys in Ontario, Canada. 
Models were applied to predict angler behavior over time at lakes included in the training phase (known lakes), as well as to predict behavior at new, unobserved lakes not used during model training (unknown lakes). 
Specifically, the following research questions were addressed:

1. How well can ML models based on \deleted{angler-generated }data \added{from an online fishing platform} predict daily angler behavior in terms of boat presence and boat counts at known lakes? 
2. Do additional data on the environment, socio-ecology and weather in the models improve the predictions?
3. Can the models be used to make predictions at unknown lakes?

The goal of the study was to advance methodological approaches in fisheries science and to demonstrate the practical utility of \replaced{data from online fishing platforms}{citizen science} in environmental monitoring\replaced{. In particular, this study aimed to assess the predictive value of angler-reported and platform-derived data for capturing spatio-temporal patterns in recreational angler effort, and to evaluate how machine-learning techniques can be used to derive actionable insights for}{, ultimately leading to} more efficient and sustainable management strategies for recreational fisheries.

\section{Materials and Methods} \label{s:methods}

\subsection{
\emph{Data}} \label{s:methods.1}
\subsubsection{Study area}
The study area comprised the province of Ontario, Canada between 2018 and 2022. 
Ontario covers more than 1 million $\mathrm{km}^2$ of land and more than 150,000 $\mathrm{km}^2$ of water. 
Ontario has a population of 14.2 million people (year 2021, Statistics Canada). 
In 2020, more than a million anglers actively fished on more than 15 million days in Ontario of which more than 750,000 were residents in Canada \added{(\cite{hunt2022selected})}. 
Walleye was the most targeted species, whereby almost a fifth of more than 50 million caught fish were harvested (\cite{hunt2022selected}).

\subsubsection{Considered lakes}

Aerial data from plane flights across lakes in Ontario were provided by the Ontario government (\cite{lester2021standardized}). 
\replaced{Angler-reported data}{Some features} were taken from the online platform Angler's Atlas (\url{www.anglersatlas.com}) and \replaced{the associated}{connected} mobile phone application MyCatch. 
\added{Lakes had to be present in both data sets and assigned 1:1 to remain in the merged data set.}
\replaced{As names and sizes of lakes could differ between the aerial data set and the online platform data set, l}{L}ocations and \added{geospatial} shapes of lakes \deleted{in the aerial data set and the online platform data set} were compared. 
\deleted{Lakes had to be present in both data sets and assigned 1:1 to remain in the merged data set.} 
\replaced{If the geospatial shape of a lake}{Moreover, shapes of lakes} differed between the data sets, \replaced{spatial overlap was assessed using the proportion of shared area. L}{ and l}akes with less than 50\%\deleted{ overlapping were removed} \added{overlap in area were excluded to ensure that only lakes representing substantially the same spatial entity across both data sources were retained}.  
The resulting data set covered 187 lakes across Ontario (Fig. \ref{fig:Waterbody_map}).

\subsubsection{Boat counts}
\added{Data from Ontario’s inland lake ecosystems collected through the Broad-scale Monitoring (BsM) program were used, which provides estimates of fishing and boating activity. 
Lakes of size between 50 and 250,000 hectares were randomly selected within spatial strata defined by Fisheries Management Zones and lake size categories. 
Boating activity data were gathered through aerial surveys conducted between 9:00 and 17:00 on randomly selected weekend, holiday, and weekday dates during the BsM cycle 3 (2018-2022) (\cite{lester2021standardized}).
See \textcite{lester2021standardized} for more details.}
\replaced{In this study, l}{L}ake-wide instantaneous angling boat counts \deleted{between 9:00 and 16:00 }were \replaced{used}{observed} with 1-31 \replaced{observation}{flights and} days per lake (15 on average) for a total of 181 \replaced{different dates}{days} between 2018 and 2022 (May-Sep) with up to 41 lake observations on a specific day (Figs. \ref{fig:count_times}, \ref{fig:fligh_freq}, \ref{fig:obslakesonday}).

The data set consisted of 2,813 samples: 1,372 samples with an absence of angling boats and 1,441 samples with presence of boats were available. 
At 10\% of the lakes, \replaced{all observation days had boat presence}{there were always boats present on the observation days} and at 24\% of the lakes, there was \replaced{no observation day}{never an angling boat observed} \added{with boat presence} (Fig. \ref{fig:Fractions_BoatAbsence}).

\replaced{The start time of a}{A}ngling boat counts did not vary much over the day\deleted{time} (\deleted{mean time 11:37am, SD 1h6min, see }Fig. \ref{fig:count_times}). 
At a specific lake, the mean number of angling boats over \replaced{all corresponding observation days}{time} was 3.4 boats (minimum 0 boats, maximum 111 boats at lake), with a standard deviation of 2.4 boats (minimum 0 boats, maximum 51 boats). 
On a specific day, the mean number of angling boats over the observed lakes on that day was 4.1 boats (minimum 0 boats, maximum 28.1 boats on a day), with a mean standard deviation of 7.5 boats (minimum 0 boats, maximum 42 boats).  
The spatial variability was therefore greater than the temporal variability.

\added{This study focused exclusively on boat-based angler effort, as shore anglers were not reliably detected through aerial surveys. As such, all analyses and predictions were limited to angler effort from boats.} 
%('AnglingTotalBoatCount')

\subsubsection{Features}
Spatial, temporal and spatio-temporal features were used as predictors in the ML models (Table \ref{tab:features}). 
The temporal resolution was daily, and the spatial resolution was on the lake level. 
The input features stayed the same for the different ML methods (Table \ref{tab:features}).

\added{Spatio-temporal features included the start time of the aerial boat count, data from the Angler's Atlas online platform and the associated MyCatch mobile phone application, fisheries management information and weather.}
\deleted{Citizen-reported angler trips were taken from the online platform Angler’s Atlas (www.anglersatlas.com) and the connected mobile phone application MyCatch.}
\deleted{The study was reviewed and approved by the Research Ethics Board of the Alberta Research Information Services (ARISE, University of Alberta), study ID \texttt{MS5\_Pro00102610}.}
\replaced{Data from the online platform and mobile phone application were divided into angler-reported data and platform-derived data. Angler-reported data included}{Features comprised} the number of reported \added{fishing} trips, the total fishing duration of the reported trips, and the mean catch rate (fish/hour) over the reported trips on a specific day at a \replaced{lake}{water body}. 
\added{Platform-derived data included the features ``Website visits'' and ``Active AA event''. 
Angler's Atlas provided informational websites for each lake on its platform which give information such as contour maps, fishing regulations and present fish species. 
Daily tracked unique website visits were saved in the feature ``Website visits'', whereby unique is defined as an individual with multiple visits was only counted once.
The feature ``Active AA event'' indicates whether Angler's Atlas conducted a competitive fishing event at the lake on that date.}
Of the 2,813 daily samples, 49 samples had reported trips (46 with one trip, 3 with two trips) and 1,614 samples had tracked ``Website visits'' of a lake information website over the last seven days, with up to 57 visits for a lake on a date.
\added{Two features on fisheries management were included as they may impact angler effort. 
The feature ``Lake closure" indicated if the lake was closed for fishing. 
The feature ``Stocking event in year" was set to ``true'' if a stocking event of hatchery fish took place, and set to ``false'' if no stocking event took place or no information was available.
}
\added{The daily weather was obtained from the simulation model BioSIM (\cite{Regniere2017}). 
See SI Methods for details.}

\added{Temporal features included the weekend day or weekday, public holiday with connected weekend, month.
``Covid-19 cases in the last seven days" on a specific date in Ontario were also considered as Covid-19 led to changes in angler behavior (\cite{Trudeau2022, Midway2021, Howarth2021, Gundelund2021MP}). 
}

\added{Spatial features comprised information on the lake environment, present fish species and humans in the surrounding area (Table \ref{tab:features}). 
The shoreline length of a lake was taken from the internal Angler's Atlas database.}
\replaced{Information on fish species, reported through catches on the online platform or mobile phone application, was}{The anglers could indicate the fish species of their catches, which were} used \replaced{to define}{for} the characteristics in the category ``Fish species'' (Table \ref{tab:features}). 
\added{Human-related features comprised the human population size and median income in the surrounding area of the lake (Table \ref{tab:features}).
See Table \ref{tab:Features_source} and SI Methods for details.}

\added{Note that the location of a lake was not included as a feature in the model but was necessary for deriving some features, such as weather conditions and distances. 
Lake location was defined by the latitude and longitude of the lake's centroid, based on its geospatial area from the online platform data set.}

\spacingset{1.5}
\begin{table}[htbp]
    \centering
    \caption{Features used in the machine-learning models. 
    \deleted{Features with \dag\ were only used in the models that also considered angler-generated data. }AA - Angler's Atlas.}
    \label{tab:features}
    \begin{tabular}{lll}
        \toprule
        \textbf{} & \textbf{Category} & \textbf{Feature} \\
        \midrule
        Spatio-temporal & Aerial survey data related (1) & Count start time \\
         & \replaced{Angler-reported data}{Angler's Atlas platform} & Number of trips\deleted{\dag} \\
         & \added{from AA (3)} & Total fishing duration\deleted{\dag} \\
         &  & Mean catch rate\deleted{\dag} \\
         & \added{Platform-derived data} & Website visits\deleted{ in last seven days}\deleted{\dag} \\
         & \added{from AA (2)} & Active AA \replaced{event}{tournament} \\
         & Fisheries management (2) & \replaced{Lake c}{C}losure \deleted{type} \\
         &  & Stocking event in year \\
         & Weather (6) & Mean air temperature \\
         &  & Total precipitation \\
         &  & Relative humidity \\
         &  & Solar radiation \\
         &  & Atmospheric pressure \\
         &  & Wind speed \\
         \midrule
        Temporal & Date-related (3) & Day type: weekend \\
         &  & Day type: holiday \\
         &  & Month \\
         & \added{Covid-19-related (1)} & \added{Covid-19 cases last seven days} \\
         \midrule
        Spatial & Lake environment (3) & Distance to urban area \\
         &  & Shoreline length \\
         &  & Maximum depth \\
         & Fish species (5) & Northern pike \added{(\textit{Esox lucius})} \\
         &  & Rainbow trout \added{(\textit{Oncorhynchus mykiss})} \\
         &  & Smallmouth bass \added{(\textit{Micropterus dolomieu})} \\
         &  & Walleye \added{(\textit{Sander vitreus})} \\
         &  & Yellow perch \added{(\textit{Perca flavescens})} \\
         & Human-related (\replaced{2}{3}) & Human population \\
         &  & Median income \\
         &  & \deleted{Covid-19 cases last seven days} \\
        \bottomrule
    \end{tabular}
\end{table}
\spacingset{2}

The selected features were not correlated to each other. 
All pairs of features had a Pearson’s correlation coefficient below abs(0.7), and were not directly correlated to the target variables (Fig. \ref{fig:corr_matrix}). 
See SI methods for additional available features removed because of correlations.

% Correlation analysis done in 
% 01_AerialData_Evaluation.py on Windows (C:\Users\julia\Documents\AnglersProject\Data\AerialDataOntario)

\subsubsection{Prediction tasks and methods}

% This was done locally on Ubuntu, using scripts: 
% - C04A_ModelTraining.py: Does model training and testing for one specific method, split, feature set
% - C04B_ModelTraining_Functions.py: Functions for C04A_ModelTraining.py

% - C04C_Compare_Model_Performances.py: Executes all the different model training and testing by calling C04A_ModelTraining.py
% - C04D_Compare_Model_Performances_Functions.py: Funtions for C04C_Compare_Model_Performances.py

% - C04F_Compare_Prediction_Tasks.py: get best performing methods on ubuntu

Two target variables were predicted by models of six different ML methods, trained with two different methods for splitting the data into a training and test set, and using three different sets of features.

The two target variables were (1) the discrete variable of the angling boat counts and (2) the boolean variable whether a boat was present or not \added{(true or false)}.
The six different ML methods \deleted{were} applied for regression (and classification) were ordinary least squares linear regression (logistic regression), support vector regression (support vector machine\deleted{s}), random forest, gradient-boosted regression trees, neural network, and k-nearest neighbors (\cite{bishop2006pattern}).
\added{The various machine learning methods possess unique strengths. Linear regression and logistic regression are characterized by their straightforward and interpretable nature. Conversely, methods such as random forest and gradient boosting are known for their accuracy and robustness. Support vector machines and neural networks are particularly adept at capturing intricate patterns, while k-nearest neighbors is esteemed for its simplicity and efficacy in specific contexts (\cite{Boateng2020Basic}).}
\added{See SI Methods for a short description of each method.}
% In addition, six Bayesian networks were learned (naive Bayes, tree-augmented naive Bayes, chow-lui for structure learning, with maximum likelihood estimation and Bayesian estimation for parameter learning, respectively) \textcolor{red}{(ref)}.  

\begin{figure}
  \includegraphics[width=\linewidth]{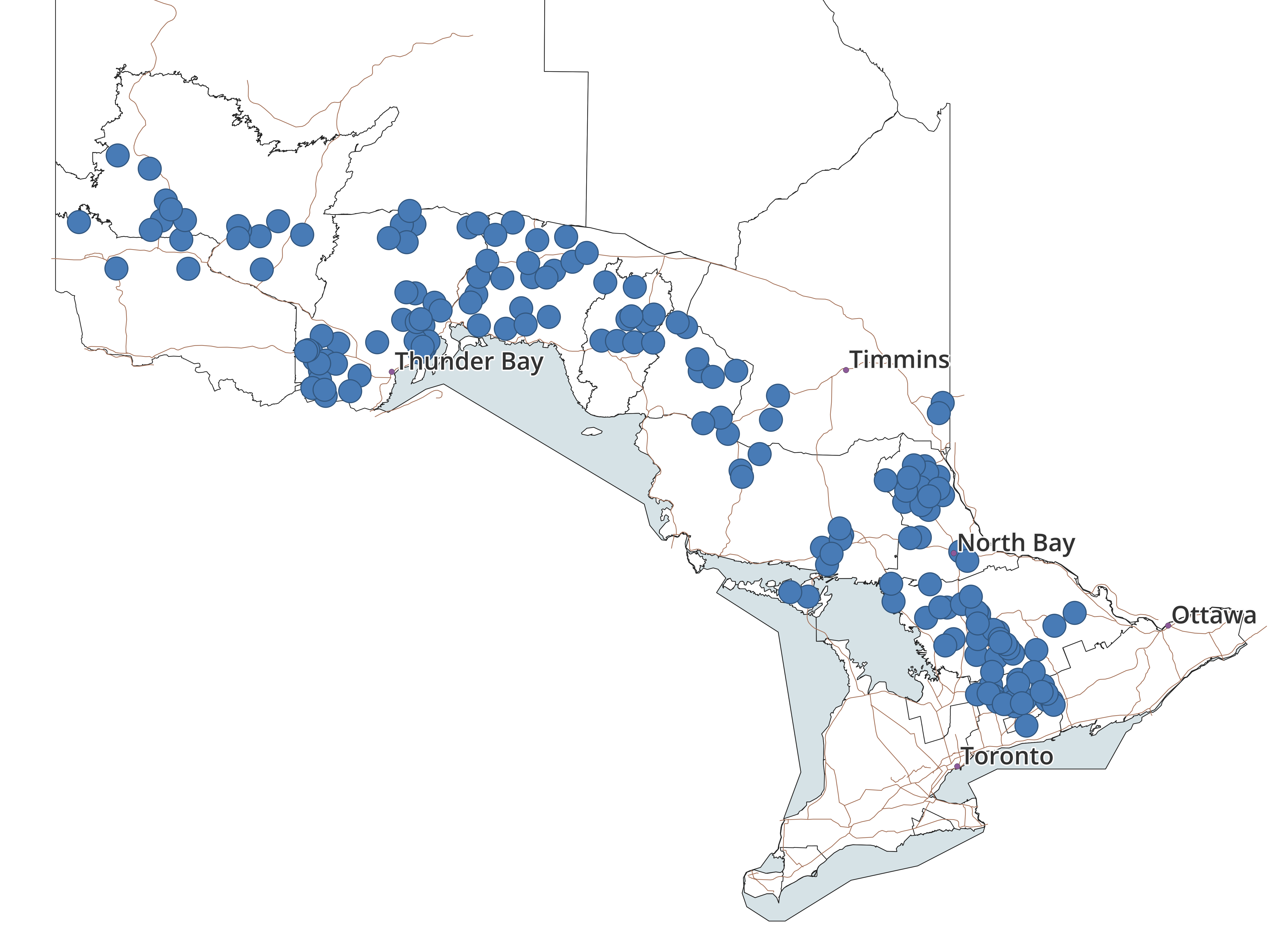}
  \caption{The 187 lakes across Ontario considered for model training and testing. Black lines show borders of different management units and brown lines indicate roads.}
  \label{fig:Waterbody_map}
\end{figure}

% C03_Data_Splits.py on Ubuntu -> creates .txt files for the lake splits that go into the model training
For splitting the data into training and test sets, the two different methods considered were: 
(1) \added{Random training-test splitting:} Division of the data set randomly into five parts and (2) \added{independent lakes splitting:} division of the data set based on lakes, which means that all measurements at a specific lake could only be in one of five parts 
(37-38 lakes with 562-563 samples, respectively, Fig. \ref{fig:Waterbody_map}).
For each ML method, five ML models were trained and tested, whereby in each model four of the five parts corresponded to the training test and the remaining part was used for model testing, respectively.  
By randomly splitting the data for training and testing (first method), the models primarily focused on predicting \replaced{angler effort}{angling pressure} at known lakes on new, unobserved days, as samples from the same lake could be present in both the training and test sets. 
In contrast, models using a split by lakes (second method) predicted \replaced{angler effort}{angler pressure} at entirely new unknown lakes and on different days, as all samples from a given lake were exclusively allocated to either the training or test set.

All prediction tasks were done three times, \replaced{each with a different}{differing in the} set of features \added{but using the same training-test splits}. 
Models were trained based on (1) only one feature, namely ``Website visits\deleted{ in the last seven days}", 
(2) with all features including the category \replaced{``Angler-reprted data from Angler's Atlas}{platform} \added{and the feature ``Website visits''} (28 features) and (3) features \replaced{excluding}{without} the category \replaced{``Angler-reprted data from Angler's Atlas}{platform} \added{and the feature ``Website visits''} (24 features, Table \ref{tab:features}). 
See SI Methods for details on ML methods with only one feature.

Model performance was evaluated on the test set\added{s} by the $\mathrm{R}^2$ values for the regression tasks, and by the accuracy \deleted{values }(percentage of correctly classified samples)\added{, Precision (proportion of correctly predicted positive cases among all predicted positives), Recall (proportion of correctly predicted positive cases among all actual positives), and F1-Score (harmonic mean of Precision and Recall)} for the classification tasks. 
\added{See SI Methods for a more detailed description of the metrics.}

\subsubsection{Feature importance \deleted{and the role of angler-generated data}}

% - C04E_Evaluate_Model_Performances.py: get best performing models on ubuntu
% - C04G_Feature_importances.py on ubuntu (runs C04A_ModelTraining.py)

In feature importance permutation, the values of a single feature in the test set were randomly shuffled, and the resulting degradation in model score (i.e., the $\mathrm{R}^2$ or accuracy value of the test set) was observed (\cite{breiman2001random}). The features were ranked according to their importance to the model, based on the extent the model performance degraded. 

Feature importance analysis was performed for the best model of each of the five training-test splits per prediction task (i.e., regression or classification) and method of training-test split (i.e., random or spatially by water bodies). As the importance of a feature refers to its information contribution to the model prediction, only ML models that had a prediction score above 0.7 ($\mathrm{R}^2$ or accuracy value of the test sets) were considered. The average importance scores for each feature were calculated over the five models of the different training-test splits. 
% For each model, the five most important features were listed and the feature's frequency was compared over all models to obtain their overall importance in the prediction task given the specific training-test split.

%In addition to the permutation feature importance, the role of citizen-reported data from AA and of the location were measured by conducting the same analysis described in the previous section without the features reported on the Angler's Atlas platform ((number of trips, webpage views, fishing durations and catch rates)) and related to location (latitude and longitude)

%To measure the role of citizen-reported data from AA: Compare to predictions of three best-performing models without features of AA angler behavior (number of trips, webpage views, fishing durations and catch rates) to estimate the value of this data source (training of new models with less features)

\section{Results} \label{s:results}

\subsection{\emph{Predictions}} \label{s:results.1}
\subsubsection{Models \replaced{using only}{with} ``Website visits"}
Using only the feature ``Website visits\deleted{ in the last seven days}", models achieved an average accuracy of 78\% for predicting angler boat presence at known lakes, based on the top-performing methods Random Forest, Gradient Boosted Regression Trees, and Support Vector Machine\deleted{s} (Table \ref{tab:pred_performance}).
For predictions at unknown lakes, the performance remained consistent at 78\% with both Random Forest and Gradient Boosted Regression Trees.
\added{Among the predictions of boat presence at known and unknown lakes, 80\% were correct and 25\% of the boat presences were missed (Table \ref{tab:add_pred_performance}).}

In contrast, the models were unable to accurately predict spatio-temporal boat counts, yielding $\mathrm{R}^2$ values of only 0.1 on both known and unknown lakes.

\subsubsection{Models \replaced{using all features, including}{with} angler-\replaced{reported}{generated} data \added{and ``Website visits''}}

Incorporating all features listed in Table \ref{tab:features} in the models slightly improved boat presence predictions to 82\% accuracy at known lakes with Gradient Boosted Regression Trees (Table \ref{tab:pred_performance}). 
\added{Precision (82\%) and recall (82\%) also increased combared to the models using only the website visits feature (Table \ref{tab:add_pred_performance}).}
At unknown lakes, the accuracy remained at 78\%, comparable to models using only the website visits feature.
\added{Precision and recall slightly decreased to 79\%, respectively (Table \ref{tab:add_pred_performance}).}

For daily boat count predictions at known lakes, the models with Gradient Boosted Regression Trees and Random Forests achieved $\mathrm{R}^2$ values of 0.8. For unknown lakes, the $\mathrm{R}^2$ values for boat count predictions dropped significantly, averaging only 0.2 with the best performing method, Support Vector Machines.

\subsubsection{Models \replaced{excluding}{without} angler-\replaced{reported}{generated} data \added{and ``Website visits''}}

Excluding angler-\replaced{reported}{generated} data \added{and the feature ``Website visits''} from the models did not reduce the accuracy of boat presence predictions at known lakes, where accuracy remained at 82\% with Random Forests and Gradient Boosted Regression Trees, nor at unknown lakes, where accuracy was 77\% with Random Forests (Table \ref{tab:pred_performance}).
\added{Moreover, precision and recall did not remarkably differ (82\% and 83\% for known lakes, 78\% and 80\% for unknown lakes, Table \ref{tab:add_pred_performance}).}

Similarly, for boat count predictions at known lakes, removing angler-\replaced{reported}{generated} data \added{and the feature ``Website visits''} did not affect model performance, which maintained $\mathrm{R}^2$ values of 0.8 with Gradient Boosted Regression Trees (Table \ref{tab:pred_performance}). At unknown lakes, the models again failed to accurately predict boat counts, with $\mathrm{R}^2$ values dropping to 0.2 on average.

%Considering solely angler-reported data in the models resulted in a 79\% accuracy in predicting the presence of boats (gradient boosted regression trees), and in a much lower R² of 0.2 for predicting the number of boats (gradient boosted regression trees, Table \ref{tab:pred_performance}).

\spacingset{1.5}
\begin{table}[htbp]
    \centering
    \begin{threeparttable}[b]
    \caption{\replaced{Performance scores of the b}{B}est performing ML methods for different prediction tasks. 
    Comparison of average \deleted{model }performance \replaced{of models using only}{ with} the feature ``Website visits\deleted{ in last seven days}'', and \replaced{models using}{with} features from Table \ref{tab:features} \replaced{including, and excluding}{with and without} angler\added{-reported data and ``Website visits''}\deleted{ (``angler data")}. 
    Spatio-temporal predictions were made at same lakes as used in model training (``known lakes''\added{, random training-test splitting}) and at lakes that were unkown for the models (``unknown lakes"\added{, independent lakes splitting}). 
    Performance scores are the $\mathrm{R}^2$ value for boat counts, and accuracy score for boat presence. 
    The mean was taken over the five models trained over different training-test data splits, respectively. Only positive mean $\mathrm{R}^2$ values were considered.}
    \label{tab:pred_performance}
    \scriptsize
    \begin{tabular}{|l|l|c|c|c|c|c|c|c|}
        \hline
        & & & \multicolumn{2}{c|}{\textbf{Only ``Website visits''}} & \multicolumn{2}{c|}{\textbf{Including angler-reported}} & \multicolumn{2}{c|}{\textbf{Excluding angler-reported}} \\
        & & & \multicolumn{2}{c|}{} & \multicolumn{2}{c|}{\textbf{data and ``Website visits''}} & \multicolumn{2}{c|}{\textbf{data and ``Website visits''}} \\ \hline
        \textbf{Target} & \textbf{Prediction} & \textbf{ML} & \textbf{Perform} & \textbf{Perform} & \textbf{Perform} & \textbf{Perform} & \textbf{Perform} & \textbf{Perform} \\
        \textbf{variable} & \textbf{task} & \textbf{method} & \textbf{train set} & \textbf{test set} & \textbf{train set} & \textbf{test set} & \textbf{train set} & \textbf{test set} \\ \hline
        \textbf{Boat} & Known & RF & 0.777 & 0.777 & 1.000 & 0.813 & 1.000 & 0.819 \\
        \textbf{presence} & lakes & GBRT & 0.777 & 0.777 & 0.876 & 0.815 & 0.876 & 0.817 \\
         & & SVM & 0.766 & 0.776 & 0.841 & 0.798 & 0.836 & 0.798 \\
         & Unknown & RF & 0.777 & 0.776 & 1.000 & 0.782 & 1.000 & 0.769 \\
         & lakes & GBRT & 0.777 & 0.776 & 0.879 & 0.780 & 0.878 & 0.768 \\
         & & \replaced{logReg}{LM} & 0.765 & 0.768 & 0.795 & 0.763 & 0.784 & 0.759 \\ \hline
        \textbf{Boat} & Known & GBRT & 0.216 & 0.135 & 0.925 & 0.754 & 0.925 & 0.773 \\
        \textbf{counts} & lakes & RF & 0.215 & 0.135 & 0.966 & 0.759 & 0.967 & 0.772 \\
         & & NN & 0.146 & 0.145 & 0.716 & 0.624 & 0.721 & 0.643 \\
         & Unknown & SV\replaced{R}{M} & 0.058 & 0.089 & 0.211 & 0.198 & 0.241 & 0.212 \\ 
         & lakes & \replaced{linReg}{LM} & 0.128 & 0.042 & 0.353 & 0.067 & 0.340 & 0.060 \\ \hline
    \end{tabular}
    \begin{tablenotes}
        \item \added{RF- Random Forest, GBRT - Gradient-Boosted Regression Trees, SVM - Support Vector Machine, LogReg - Logistic Regression, NN - Neural Network, SVR - Support Vector Regression, LinReg - Linear Regression}
    \end{tablenotes}
   \end{threeparttable}
\end{table}
\spacingset{2}

\subsection{\emph{Feature importance}} \label{s:results.2}

\subsubsection{Models \replaced{using all features, including}{with} angler-\replaced{reported}{generated} data \added{and ``Website visits''}}
For predicting the presence or absence of boats at known and unknown lakes, the feature ``Website visits'' was the most important feature with roughly twice to three times as much influence on prediction performance as the second most important feature (Table \ref{tab:feature_importance}). 
Other important features were the distance to an urban area and the shoreline length. 
Besides these, information on the presence of fish species, namely walleye, smallmouth bass and yellow perch contributed most information to the model predictions.

Predictions of boat counts at unknown lakes resulted in performance scores below 0.7 ($\mathrm{R}^2$ and accuracy) and were, hence, not considered in the feature importance analysis (Table \ref{tab:feature_importance}).

\subsubsection{Models \replaced{excluding}{without} angler-\replaced{reported}{generated} data \added{and ``Website visits''}}
For predicting the presence or absence of boats at known or unknown lakes, the importance of features in ML models without angler-\replaced{reported}{generated} data \added{and ``Website visits''} did not differ from the models with angler-\replaced{reported}{generated} data \added{and ``Website visits''}. 
Instead of the feature ``Website visits'', atmospheric pressure and the presence of smallmouth bass belonged to the five most important features, respectively. 

For the temporal prediction of boat counts at known lakes, feature permutation revealed that again shoreline length and distance of the lake from an urban area were important predictors. 
The other important features were human population in the area, day type (weekend), and wind speed (Table \ref{tab:feature_importance}).

\spacingset{1.5}
\begin{table}[htbp]
    \centering
    \caption{Comparison of feature importance in models \replaced{including and excluding}{with and without} angler-\replaced{reported}{generated} data \added{and website visits (WV)} from the \replaced{online fishing}{Angler's Atlas} platform. Importance shows the extent the model performance degraded ($\mathrm{R}^2$ value or accuracy value) when the values of the feature were randomly shuffled.}
    \label{tab:feature_importance}
    \small
    \begin{tabular}{|l|c|ll|ll|}
        \hline
        & & \multicolumn{2}{c|}{\textbf{Models including angler-reported}} & \multicolumn{2}{c|}{\textbf{Models excluding angler-reported}} \\
        & & \multicolumn{2}{c|}{\textbf{data and ``Website visits''}} & \multicolumn{2}{c|}{\textbf{data and ``Website visits''}} \\
        & \textbf{Rank} & \textbf{Feature} & \textbf{Importance} & \textbf{Feature} & \textbf{Importance} \\ \hline
        \textbf{Boat present} & 1 & Website visits & 0.076 $\pm$ 0.011 & Distance to urban area & 0.031 $\pm$ 0.008 \\
        Known lakes & 2 & Distance to urban area & 0.028 $\pm$ 0.009 & Shoreline length & 0.028 $\pm$ 0.007 \\
        & 3 & Shoreline length & 0.020 $\pm$ 0.007 & Smallmouth bass & 0.022 $\pm$ 0.007 \\
        & 4 & Walleye & 0.016 $\pm$ 0.005 & Walleye & 0.021 $\pm$ 0.007 \\
        & 5 & Smallmouth bass & 0.015 $\pm$ 0.006 & Atmospheric pressure & 0.016 $\pm$ 0.005 \\ \hline
        \textbf{Boat present} & 1 & Website visits & 0.077 $\pm$ 0.013 & Shoreline length & 0.047 $\pm$ 0.010 \\
        Unknown & 2 & Walleye & 0.043 $\pm$ 0.007 & Smallmouth bass & 0.040 $\pm$ 0.009 \\
        lakes & 3 & Shoreline length & 0.033 $\pm$ 0.008 & Walleye & 0.039 $\pm$ 0.008 \\
        & 4 & Yellow perch & 0.032 $\pm$ 0.008 & Distance to urban area & 0.029 $\pm$ 0.009 \\
        & 5 & Distance to urban area & 0.027 $\pm$ 0.009 & Yellow perch & 0.021 $\pm$ 0.006 \\ \hline
        \textbf{Boat Count} & 1 & Distance to urban area & 0.338 $\pm$ 0.067 & Shoreline length & 0.422 $\pm$ 0.048 \\
        Known lakes & 2 & Shoreline length & 0.334 $\pm$ 0.040 & Distance to urban area & 0.334 $\pm$ 0.060 \\
        & 3 & Human population & 0.223 $\pm$ 0.026 & Human population & 0.211 $\pm$ 0.027 \\
        & 4 & Day type (weekend) & 0.095 $\pm$ 0.026 & Day type (weekend) & 0.101 $\pm$ 0.025 \\
        & 5 & Wind speed & 0.066 $\pm$ 0.022 & Wind speed & 0.068 $\pm$ 0.029 \\ \hline
    \end{tabular}
\end{table}
\spacingset{2}
	
\section{Discussion} \label{s:discussion}

%Research question: 1.	To what extent can machine-learning models based on environmental and human data predict angler pressure in terms of boat counts? 2. Do angler-generated data from an online platform improve these predictions? 3.	What features are most important for the predictions?

% Key messages:
% 1. Website visits (angler-generated data) are sufficient to predict the presence or absence of boats on a lake on a specific day, but poor predictors for number of boats.
% 2. Environmental, socio-ecological, and weather variables together with ML methods enable to predict boat counts on lakes. 
% 3. Predictions at unknwon lakes are not possible.

\added{This study examined the utility of spatio-temporal, temporal and spatial features to predict recreational angler effort across lakes in Ontario, Canada, testing a variety of machine-learning methods.}
\replaced{Importantly, the utility of data from an online fishing platform and its associated mobile application was shown, and in particular}{Angler-generated data, particularly} recent lake website visits \deleted{,} were effective in predicting the presence or absence of \added{angling} boats on a given day but were insufficient for accurately predicting boat counts at the lakes. 
In ML models predicting daily boat counts at known lakes, where\deleted{by} up to 77\% of the \deleted{boat count }variance could be explained\deleted{ by the model}, the most important features were shoreline length, distance to urban areas, and human population. 
\replaced{However, these models failed to generalize to unknown lakes across Ontario due to the limited predictive power of available features and data.}{The available data and features were insufficient to accurately predict boat counts at unknown lakes across Ontario.}

Website visits emerged as the most important feature for predicting boat presence \replaced{across}{at} both known lakes over time and \deleted{at} entirely unknown lakes. 
\replaced{These}{Website} visits reflect information exchange and user engagement on the online platform, serving as a proxy for angler \replaced{interest and intent to fish}{activity}.
\added{The utility of website visits was also supported in a previous study, where a Bayesian network found a direct relationship between website visits, boat counts, and fishing duration (\cite{taheri2025webpage}).}  
\replaced{Angler-reported}{Other angler-generated} features \replaced{from the online platform had limited}{showed only minor} importance, likely due to \replaced{data sparsity and variable reliability}{limited data availability}.
\deleted{Previous studies have explored the value of angler-generated data in various contexts. 
For example, online platforms have been used to monitor angler satisfaction and identify key predictors by applying advanced modeling techniques (\cite{vacura2023makes, gundelund2022investigating}). 
Additionally, comparisons with traditional surveys of fishing activity revealed many similarities (\cite{johnston2022comparative}). 
Furthermore, changes in fishing activity due to Covid-19 have been analyzed using data from online platforms (\cite{gundelund2021changes}).}

\replaced{This study builds upon earlier efforts to model angler effort using diverse datasets and methodologies (\cite{jensen2022phenomenological, askey2018angler, powers2016estimating, hunt2019predicting}).} 
{Models for predicting angler effort were also explored in previous studies, achieving similar accuracy with different features and methods (\cite{jensen2022phenomenological, askey2018angler, powers2016estimating, hunt2019predicting}).}
For instance, \replaced{\textcite{jensen2022phenomenological} used an autoregressive Poisson model with}{a study predicting daily boating effort at the mouth of the Columbia River, USA, reached Pearson $\mathrm{R}^2$ values of up to 79\% using an overdispersed Poisson likelihood-based generalized linear model with an autoregressive structure, and} creel survey data\added{ to predict daily boating effort at the Columbia River, achieving Pearson $\mathrm{R}^2$ values up to 79\%.}\deleted{(\cite{jensen2022phenomenological}).} 
\replaced{Similarly, \textcite{askey2018angler} combines a}{A} generalized linear mixed model \replaced{with time-lapse camera data to estimate annual boat counts across lakes in British Columbia, reaching an $\mathrm{R}^2$ value of 0.68.}{combined with the Small Lakes Index Management sampling protocol yielded an $\mathrm{R}^2$ value of 0.68 for annual boat counts across lakes in British Columbia, Canada, based on time-lapse camera data (\cite{askey2018angler})}.
Bayesian methods \replaced{also proved effective for predicting seasonal angler effort,}{were applied to predict seasonal angler pressure} using drone and fishing app data in a Lithuanian reservoir, and \deleted{to predict fishing effort per day type using different} creel data and aerial surveys for a winter fishery in Ontario (\cite{dainys2022angling, tucker2024estimating}). 
\replaced{The present study expands this work demonstrating the predictive}{Here, The} value of \deleted{angler-generated data } from an online \added{fishing} platform\deleted{, in comparison to commonly used features} in \replaced{machine-learning}{prediction} models\deleted{, was demonstrated}.
%The models often focus on temporal predictions and disregard the attempt to get higher resolution over space.
%These models tend to focus primarily on temporal predictions, often at the expense of higher spatial resolution. The here presented approach incorporates both temporal and spatial elements to predict angler behavior across multiple lakes.
% R2 of 0.71 for red snapper angler per hour using dockside cameras \cite{powers2016estimating} 

\replaced{Among the most influental predictors across}{Important features for} all models \deleted{and prediction tasks }were the shoreline length and the distance to urban areas\deleted{ of a lake}, although \added{they were} not directly correlated \replaced{with daily}{to the} boat counts. 
The shoreline length affects \replaced{angling opportunity and access points}{the availability of fishing spots and boat access}, while the distance to urban areas affects \replaced{accessibility and angler convenience}{how easily angler can get to the lake}. 
\replaced{H}{In addition, h}uman population \replaced{,}{and} day type \replaced{(e.g., weekends or holidays) and the occurrence of certain fish species also played important roles}{were relevant for predicting boat counts, and the occurrence of certain fish species for the presence of boats}. 
The human population in the surrounding area affects the overall demand for recreational fishing. 
\deleted{Human population size was also used in a logistic population growth rate model to spatially extrapolate angler density utilizing an angler participation rate relationship (\cite{post2008angler}).}
Weekends and holidays \replaced{typically exhibit higher angler effort, consistent with prior studies}{generally see higher recreational activity as more people have time off work. 
A higher angler pressure during weekends was also detected in other studies and significant predictors in their models} (\cite{van2015imputing, jensen2022phenomenological, hunt2007predicting, dainys2022angling, askey2018angler}).
Wind speed and atmospheric pressure \added{also} emerged as \replaced{relevant, reflecting the physical and behavioral constraints on both boating and fish activity}{important features in predicting fishing activity. 
High wind speeds can make boating and fishing on lakes both dangerous and uncomfortable, while also influencing fishing success} (\cite{kuparinen2010abiotic, stoner2004effects}). 
\deleted{Atmospheric pressure affects fish behavior by influencing their activity and distribution patterns, which in turn can impact fishing catch rates (\cite{stoner2004effects}).} 
%Both features were also used in the model by (\cite{jensen2022phenomenological}).

Random forests \deleted{(RF)} and gradient-boosted regression trees \replaced{consistently outperformed other ML models, which aligns with previous findings in spatio-temporal prediction tasks}{(GBRT) were the top-performing machine learning models overall, consistent with other studies comparing model performance for spatio-temporal predictions} (\cite{kim2022machine, ahmad2017trees}). 
\replaced{T}{One reason for t}heir \replaced{better}{superior} performance may \replaced{stem from their ability to capture}{be that they handle} complex, non-linear relationships \replaced{and handle heterogeneous data}{well} without \deleted{requiring }extensive \deleted{hyperparameter }tuning. 
In contrast, \deleted{models like }neural networks and support vector machines \replaced{typically require intensive}{(SVMs) often depend more heavily on careful} hyperparameter \replaced{optimization}{tuning} to reach their full potential, \replaced{a step not undertaken}{which was not performed} in this analysis (\cite{taylor2021sensitivity, weerts2020importance, mantovani2015tune}).

%Another reason could be that no hyperparameter tuning was performed in this study, but default values in the scikit-learn Python library were used for all models. 
%Limitations regarding features (same features used for all ML methods, but some methods benefit from more / less features)

\deleted{The temporal resolution of the dataset was limited, with an average of only 36.2 unique days per open-water season across all lakes over five years. 
This may have hindered the detection of actual existing relationships between variables, such as website visits and boat counts. 
Most of the key features for predicting boat counts showed only spatial variation, likely reflecting the greater spatial variation in boat counts compared to the temporal variation within the dataset.} %\textcolor{red}{(temporal only day type, wind speed and atmospheric pressure)}

\added{Limitations of smartphone application and online platform data for temporal predictions were evident in this study. 
Although website visits were among the most important predictor for boat presence across lakes, the current data and modeling approach did not support reliable detection of temporal changes in boating activity. 
The temporal resolution of the dataset was limited, and while spatial predictors like human population density or shoreline length explained variation among lakes, angler-reported data and website visits did not clearly capture dynamic responses to changing conditions such as weather or fish activity. 
This points to a key limitation of using such data for forecasting angler behavior over time, despite the common expectation that real-time digital platforms could offer this kind of insight. 
Capturing adaptive responses would likely require finer-resolution data or different modeling frameworks that can identify deviations from baseline patterns.}

\added{This study is limited to modeling boat-based angler effort, as aerial survey data captured only boats on the lakes. 
However, the angler-reported data used in the models may include both boat and shore-based anglers, which introduces some uncertainty in matching predictor and response variables. 
In regions like Ontario, shore-based angling can account for a substantial portion of total effort. 
Accordingly, this mismatch should be acknowledged when interpreting results and comparing them to other angling data sources that more comprehensively capture shore-based effort.}

\replaced{Future work could expand the spatial and temporal scope of the models, potentially improving predictive power.}{A possible extension of the study is to use the models and data for predictions at different scales.} 
For example, \replaced{incorporating data at broader spatial or temporal scales — such as regional angler effort or year-to-year variation —}{predicting broader trends such as regional angler effort or annual variations} could enhance \replaced{long-term forecasts}{model accuracy}.
\replaced{Alternatively, d}{D}ata \deleted{samples could also focus }on individual anglers \replaced{could be used to predict angler decisions based on personal attributes, such as}{rather than a single day at a water body, enabling predictions of angler choices based on factors like} socioeconomic background (\cite{kaemingk2020harvest,schmid2025analyzing}). 
\added{In this context, incorporating the degree of angler specialization, as conceptualized in recreational specialization theory, could offer further explanatory power for understanding angler preferences and decision-making. 
This framework, which categorizes anglers based on their skill level, commitment, and behavioral patterns, has been widely used to explain variability in angling behavior (\cite{Karpiński2021Environmental, Beardmore2013Evaluating, Salz2001Development}).}
\replaced{As seen in other studies, integrating angler-reported}{Angler-generated} data from online platforms \replaced{offers a scalable avenue for modeling fishing activity}{can further improve model predictions of angler traffic, as seen in previous studies} (\cite{fischer2023boosting}).

% Role of conventional data and different data for validation in future studies (type and regions, and ice-fishing)
Conventional surveys \replaced{remain critical}{are essential for the prediction model, as they provide a baseline} for training and validat\replaced{ing predictive models}{ion}. 
\replaced{Using}{To get more general results on the importance and prediction power of features regarding angler pressure,} other conventional data, e.g., received through creel surveys, \replaced{as ground truth data could help benchmark the predictive utility of novel features}{can be used as the target variable in future studies} (\cite{jensen2022phenomenological, pope2017estimating}). 
\replaced{Likewise, time-lapse camera and drone data can offer high-resolution records of angler effort}{Moreover, angler pressure detected by remotely operated cameras and drones can be used to train and validate the ML models} (\cite{smallwood2012expanding, morrow2022improving, van2015imputing, provost2020assessing, dainys2022angling, askey2018angler}).

\replaced{Finally, i}{I}ncorporating additional features related to habitat quality, such as water temperature\replaced{,}{ or} fish stock size, \deleted{as well as factors influencing the overall fishing experience, like access to }boat ramp\replaced{ availability and}{s or available} campgrounds \added{access}, could further enhance model \replaced{accuracy, particularly for}{predictions or even enable the prediction of boat counts at} unknown lakes (\cite{aprahamian2010examining, fischer2023boosting}). 
\replaced{While}{However,} these \replaced{variables}{features} are often difficult to \replaced{collect systematically, especially across many water bodies,  their integration alongside more readily available data could enhance both explanatory and predictive capabilities of future models}{measure or not readily available}. 
\deleted{For instance, collecting consistent fish abundance data across multiple lakes demands substantial effort, and habitat quality metrics frequently rely on localized surveys. 
Additionally, some features fluctuate over time, limiting their utility for future predictions. 
Nevertheless, integrating these factors into machine learning models, alongside more accessible data, has the potential to deepen our understanding of \replaced{recreational angler effort}{fishing activity} and angler preferences.}

% -------- Removed: --------
%For the prediction of boat counts, the five most important features over the best-performing models for each training-test split were shoreline length, distance to urban area, the human population, is weekend and the atmospheric pressure (Table \ref{}. When adding features from the anglers online platform, the five most important features stayed the same.

%For the prediction of boat present or not, the features on present fish species (walleye, yellow perch and smallmouth bass) were important besides the shoreline and distance to urban area. Important features changed when the anglers online platform features were included in the model. Besides walleye present, distance to urban area, shoreline and atmospheric pressure, the website visits in the last seven days was the second most important feature over the considered ML models.

%\section{\deleted{Conclusion}}\label{s:conclusion}
%The study demonstrated the value of angler-generated data for spatio-temporal predictions of daily lake-based recreational angler pressure. 
\deleted{The study shows the value of integrating angler data from online platforms into predictive models and highlights the potential of machine learning models to improve fisheries management and conservation strategies.}

\if0\blind{
\section{Acknowledgements}
We acknowledge the data contribution by Ontario Ministry of Natural Resources and Forestry.
We thank Dak de Kerckhove from the Ontario Ministry of Natural Resources for valuable discussions.

The study was reviewed and approved by the Research Ethics Board of the Alberta Research Information Services (ARISE, University of Alberta), study ID \texttt{MS5\_Pro00102610}.
} \fi

\section{Conflict of Interest Statement}
The authors declare no conflict of interest.

\printbibliography

\newpage
\renewcommand{\thesection}{S\arabic{section}}
\renewcommand{\thefigure}{S\arabic{figure}}
\renewcommand{\thetable}{S\arabic{table}}
\setcounter{section}{0}
\setcounter{figure}{0}
\setcounter{table}{0}

\thispagestyle{empty}  

{\bf \Large Supplementary Information: \newline Website visits are enough to predict angler presence using machine learning}
\newpage

\section{\emph{SI methods}} \label{s:SI_methods}

\subsection{\added{Features on weather from BioSIM}}
\added{We used the software tool BioSIM 11 to receive daily weather data and the elevation of each lake (\cite{Regniere2017}, Table \ref{tab:Features_source}). 
BioSIM selected the four nearest weather stations for each lake (based on the centroid of the lake) for interpolations and adjusted weather data for differences in elevation, latitude and longitude. 
Historical daily weather observations were used (Open Topo Data API Nasa srtm 30 m) and the bi-linear interpolation method was applied in the observation-based Climatic Daily model.}

\subsection{\added{Distance of a lake to the next urban area and to a road}}
\added{City boundaries and roadways were taken from Statistics Cananada (Spatial information products: Boundary files, 2021 and Road network files, 2022).} % https://www12.statcan.gc.ca/census-recensement/2011/geo/RNF-FRR/index-eng.cfm
%\added{The geometry of lakes were simplified by 0.0005 degrees using the Douglas-Peucker algorithm implemented in ''ST\textunderscore Simplify()" in PostGIS (\url{https://postgis.net}).} %method described here: https://postgis.net/docs/ST_Simplify.html
\added{The minimal Cartesian distance between the nearest points for each simplified lake and road or the centroid of a city based on their coordinates was determined using (''ST\textunderscore Distance" in PostGIS).} %https://postgis.net/docs/ST_Distance.html

\subsection{\added{Surrounding area of a lake for demographic data}}
\added{Demographic data was calculated for each lake by considering cities in the surrounding area at different distances. 
A weighted mean of the human population size, and mean and median income in the surrounding area of a lake was computed by considering three different distances (0.6 * 11 km distance + 0.3 * 111 km distance + 0.1 * 555 km distance).}

\subsection{Correlations between additional features}
\begin{figure}[ht]
  \includegraphics[width=\linewidth]{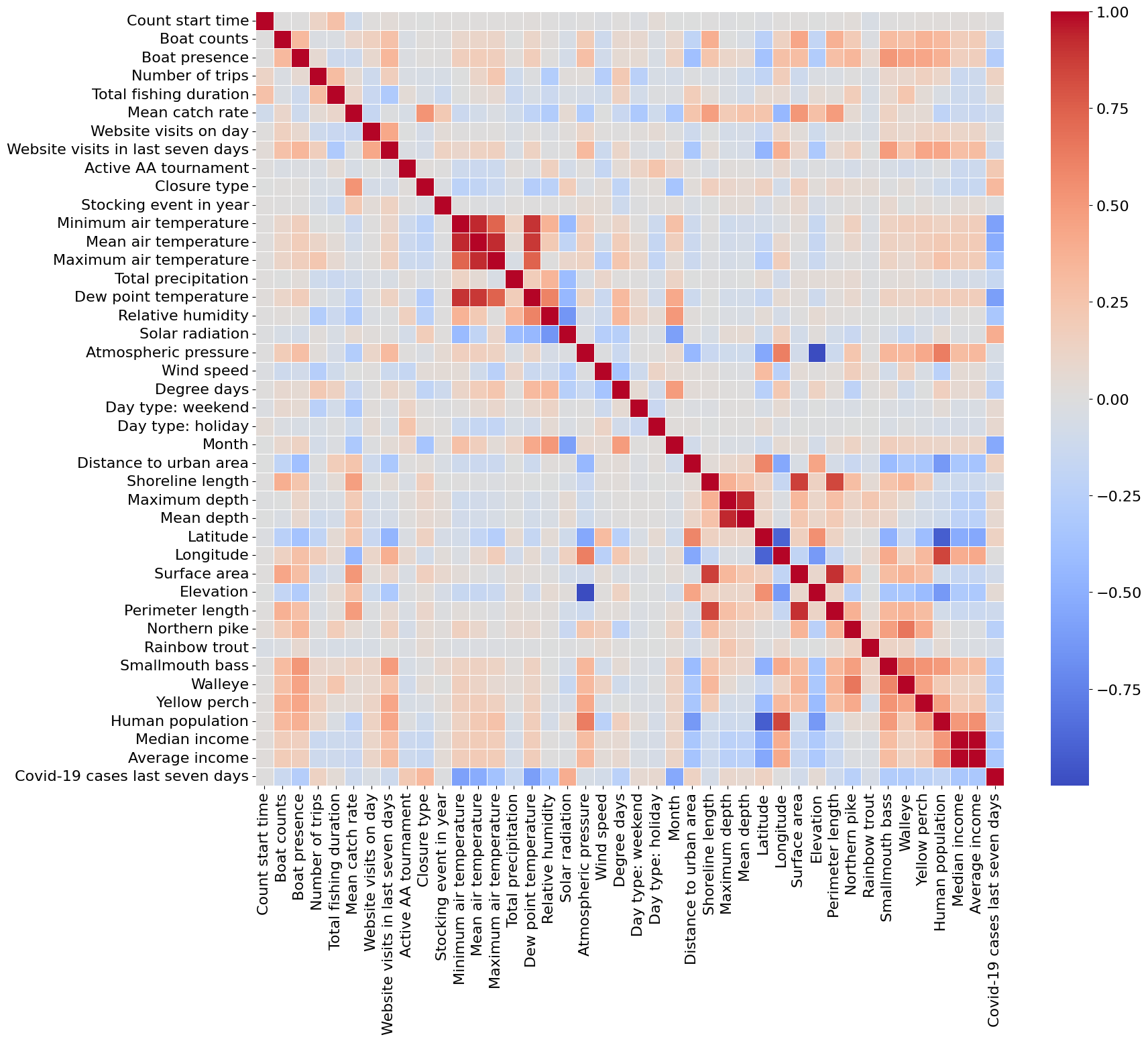}
  \caption{Pearson correlation coefficients between available variables.}
  \label{fig:corr_matrix}
\end{figure}

Additional features were available, but not considered in the models. 
Because of high correlations (absolsute Pearson correlation coefficient above 0.7), the following variables were removed:
\begin{itemize}
    \item Mean income (correlated to median income)
    \item Latitude (correlated to longitude and human population) 
    \item Longitude (correlated to latitude and human population)
    \item Minimum air temperature (correlated to mean air temperature and maximum air temperature)
    \item Maximum air temperature (correlated to mean temperature and minimum air temperature)
    \item Dew point temperature (correlated to minimum, maximum and mean air temperature)
    \item Lake surface area (correlated to shoreline length)
    \item Mean depth of lake (correlated to maxim depth of lake)
    \item Elevation of the lake (correlated to atmospheric pressure)
\end{itemize}
See Figure \ref{fig:corr_matrix} for more details.

\subsection{Machine learning methods}

\subsubsection{Ordinary Least Squares Linear Regression (Logistic Regression):}
Ordinary Least Squares (OLS) is a linear regression method that minimizes the sum of the squares of the differences between observed and predicted values. 
It is widely used for its simplicity and interpretability (\cite{Cosenza2020Comparison, Delgado2019An}).
Logistic Regression is a classification algorithm that models the probability of a binary outcome based on one or more predictor variables. 
It uses a logistic function to model a binary dependent variable (\cite{Chaurasia2020Applications}).

\subsubsection{Support Vector Regression (Support Vector Machine):}
Support Vector Regression (SVR) is an extension of support vector machines (SVM) for regression problems. 
It aims to find a function that deviates from the actual observed values by a value no greater than a specified margin (\cite{Modaresi2018A, Wu2020Do}).
Support Vector Machine (SVM) is primarily used for classification tasks and works by finding the hyperplane that best separates the data into different classes (\cite{Boateng2020Basic, Santos2021A}).

\subsubsection{Random Forest (RF):}
Random Forest is an ensemble learning method that constructs multiple decision trees during training and outputs the mode of their predictions for classification or mean prediction for regression. 
It is known for its robustness and ability to handle large datasets (\cite{Cosenza2020Comparison, Ao2019The, Santos2021A}).

\subsubsection{Gradient-Boosted Regression Trees (GBRT):}
Gradient Boosting is an ensemble technique that builds models sequentially, with each new model attempting to correct the errors made by the previous ones. 
It is effective for both regression and classification tasks and is known for its high accuracy (\cite{Delgado2019An, Chaurasia2020Applications}).

\subsubsection{Neural Network (NN):}
Neural Networks are computational models inspired by the human brain, consisting of interconnected groups of nodes (neurons). 
They are capable of capturing complex patterns in data and are used for both regression and classification tasks2 (\cite{Boateng2020Basic}).

\subsubsection{K-Nearest Neighbors (KNN):}
K-Nearest Neighbors (KNN) is a simple, non-parametric algorithm used for classification and regression. 
It predicts the value of a point based on the values of its k-nearest neighbors in the feature space. 
It is easy to implement but can be computationally expensive with large datasets (\cite{Cosenza2020Comparison, Wu2020Do}).

\subsection{Machine learning methods with one feature}

When ML models are applied to datasets with only one feature, the behavior of the models simplifies but still follows the principles of their respective algorithms.
Logistic regression fits a linear decision boundary, SVM creates a nonlinear boundary using kernel functions, and ensemble methods like random forests and gradient boosting aggregate predictions from multiple decision trees. 
KNN relies on the distance to neighboring points for classification, while MLP uses a neural network to capture more complex relationships in the data.

\subsubsection{Logistic Regression:}

Logistic regression is a linear model used for binary classification. 
With a single feature $x$, the decision function is given by:

\begin{equation}
P(y=1|x) = \frac{1}{1 + e^{-(\beta_0 + \beta_1 x)}}
\end{equation}

where $\beta_0$ is the intercept, and $\beta_1$ is the coefficient for the feature $x$. 
The model predicts class $y = 1$ if $P(y=1|x) > 0.5$, and class $y = 0$ otherwise.

\subsubsection{Support Vector Machine (SVM):}

For a single feature $x$, an SVM with the radial basis function (RBF) kernel classifies data using the decision function:

\begin{equation}
f(x) = \sum_{i=1}^{n} \alpha_i y_i \exp(-\gamma (x_i - x)^2) + b
\end{equation}

Here, $\alpha_i$ are the support vector coefficients, $y_i$ are the class labels of the support vectors, $\gamma$ controls the width of the RBF kernel, and $b$ is the bias term. The model classifies the input as class $+1$ if $f(x) > 0$, and class $-1$ otherwise.

\subsubsection{Random Forest Classifier:}

A random forest classifier with one feature constructs an ensemble of decision trees. 
Each tree splits the data based on a threshold on the feature $x$. 
The final prediction is made by averaging the predictions from all trees:

\begin{equation}
\hat{y} = \frac{1}{N} \sum_{i=1}^{N} h_i(x)
\end{equation}

where $h_i(x)$ is the prediction from the $i$-th tree, and $N$ is the number of trees in the forest. 
The majority vote (for classification) determines the final predicted class.

\subsubsection{Gradient Boosting Classifier:}

Gradient boosting with one feature works by sequentially fitting decision trees to the residual errors of the previous trees. 
The prediction for a new input $x$ is given by:

\begin{equation}
\hat{y} = \sum_{m=1}^{M} \nu \cdot h_m(x)
\end{equation}

where $M$ is the total number of trees, $h_m(x)$ is the prediction from the $m$-th tree, and $\nu$ is the learning rate that controls the contribution of each tree. 
The final class prediction is based on the cumulative sum of the individual tree outputs.

\subsubsection{K-Nearest Neighbors Classifier (KNN):}

With one feature, the K-nearest neighbors (KNN) classifier classifies a data point based on the majority label of its nearest neighbors in the feature space. 
The distance metric is typically the Euclidean distance (with $p=2$ in the Minkowski distance formula):

\begin{equation}
d(x_i, x_j) = |x_i - x_j|
\end{equation}

The KNN model predicts the class that appears most frequently among the $k$ nearest neighbors. 
If the number of neighbors is odd, the decision is made by a majority vote.

\subsubsection{Neural Network Classifier (MLP):}

The multi-layer perceptron (MLP) classifier uses a neural network for classification. 
With one feature, the input layer has one node, followed by one or more hidden layers. 
For a hidden layer of $h$ units with ReLU activation, the transformation for input $x$ is:

\begin{equation}
z_j = \max(0, w_j x + b_j), \quad \forall j \in [1, h]
\end{equation}

The output of the hidden layer is then passed through subsequent layers until the final output layer, which classifies the input as either class $+1$ or $-1$.

\subsection{\added{Additional performance metrics}}

\added{Precision, recall, and F1-score were computed to provide insights beyond the overall accuracy. 
Precision measures the proportion of correctly predicted boat presences (or absences, true postives (TP)) among all instances classified as such, quantifying the model’s ability to avoid false positives (FP):
\begin{equation}
  \text{Precision} = \frac{\text{TP}}{\text{TP} + \text{FP}}
\end{equation}}

\added{Recall, also known as sensitivity, evaluates the model’s ability to correctly identify actual boat presences (or absences) by calculating the proportion of true positives among all actual positive cases (true positives and false negatives (FN)):
\begin{equation}
  \text{Recall} = \frac{\text{TP}}{\text{TP} + \text{FN}}
\end{equation}
A high recall indicates that few boat presences (or absences) are missed.} 

\added{F1-score is the harmonic mean of precision and recall, offering a single measure that balances both metrics:
\begin{equation}
  F_1 = 2 \times \frac{\text{Precision} \times \text{Recall}}{\text{Precision} + \text{Recall}}
\end{equation} 
It is particularly useful when false positives and false negatives have comparable importance.}

\newpage
\section{SI Figures}
\begin{figure}[ht]
  \includegraphics[width=\linewidth]{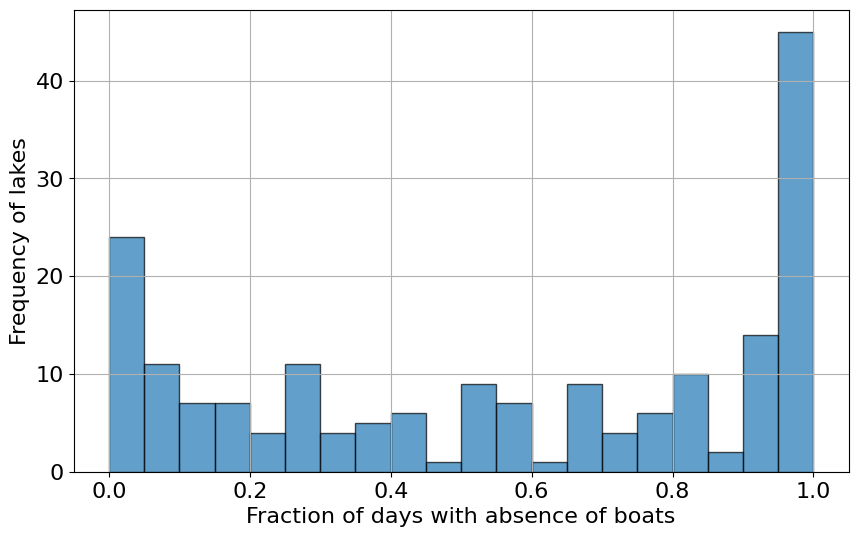}
  \caption{Frequency of fractions of absence of angling boats on observation days at the 187 lakes. At 45 lakes, there were no fishing boats detected over all observation days, and at 18 lakes, there were always boats present on the observation days.}
  \label{fig:Fractions_BoatAbsence}
\end{figure}

\begin{figure}[ht]
  \includegraphics[width=\linewidth]{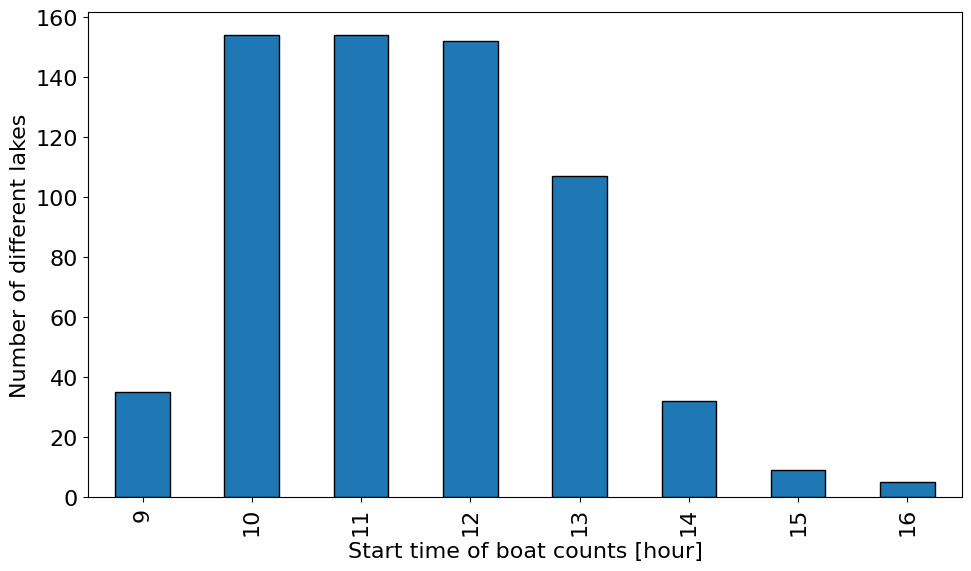}
  \caption{Start times of angling boat counts. All counts started between 9:00 and 16:00.}
  \label{fig:count_times}
\end{figure}

\begin{figure}[ht]
  \includegraphics[width=0.8\linewidth]{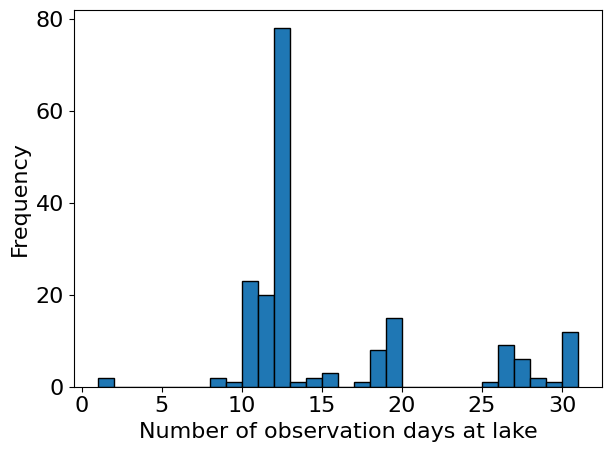}
  \caption{Frequencies of number of flights (observation days) at the 187 lakes. 13 flights was the most likely number of flights for a lake.}
  \label{fig:fligh_freq}
\end{figure}

\begin{figure}[ht]
  \includegraphics[width=0.8\linewidth]{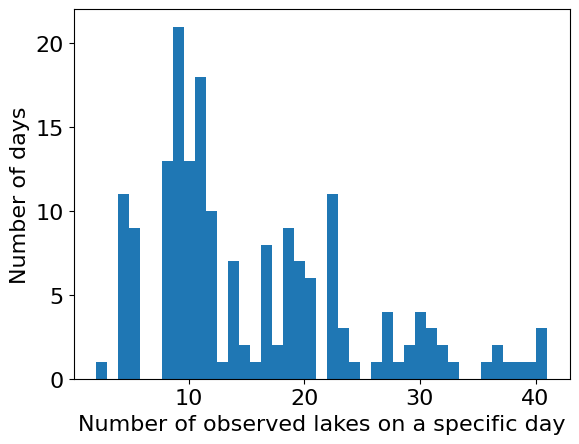}
  \caption{Frequencies of number of flights (observation days) at the lakes. Up to 41 lakes were observed on a specific day (three days). On most dates, nine lakes were observed (21 days).}
  \label{fig:obslakesonday}
\end{figure}

\begin{figure}[ht]
  \includegraphics[width=\linewidth]{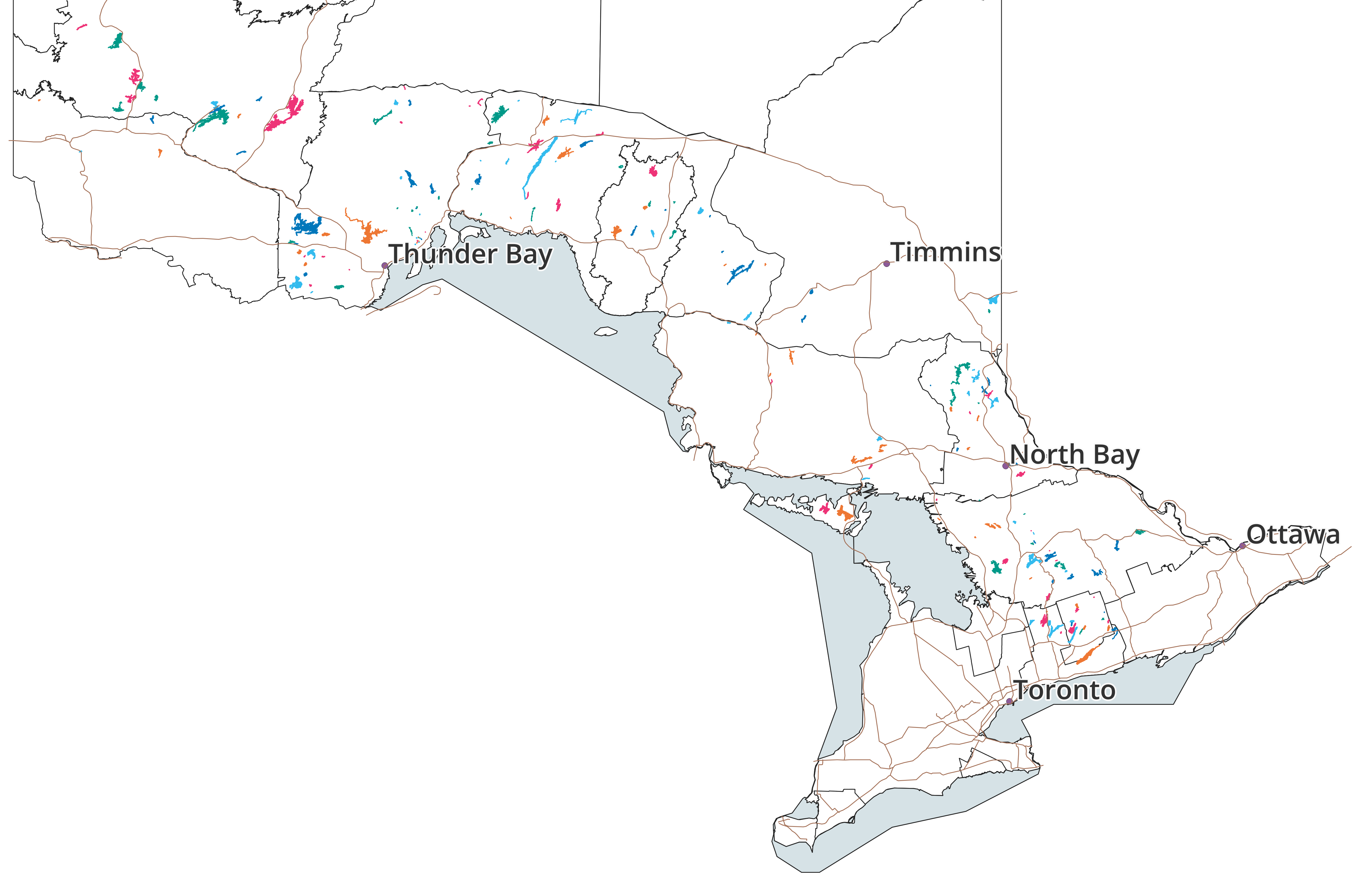}
  \caption{The 187 lakes across Ontario considered for model training and testing. The colors show the five parts that were used for training and testing the machine learning models with the division of the data set based on lakes.}
  \label{fig:Waterbody_map_splits}
\end{figure} 

\clearpage

\newpage
\section{\added{SI Tables}}
\begin{longtable}{|p{4.6cm}|p{2.0cm}|p{2.8cm}|p{4.4cm}|}
%\captionsetup{justification=raggedright,singlelinecheck=false}
\caption{\label{tab:Features_source} \added{Features used for predicting the target variables. AA - Angler's Atlas database, StatCan - Statistics Canada}}\\
\hline
\textbf{Feature} & \textbf{Data type} & \textbf{Dimensionality} & \textbf{Source} \\ \hline
\endfirsthead
\multicolumn{4}{c}{{\tablename\ \thetable{} -- continued from previous page}} \\
\hline
\textbf{Feature} & \textbf{Data Type} & \textbf{Dimensionality} & \textbf{Source} \\ \hline
\endhead
\hline \multicolumn{4}{r}{{Continued on next page}} \\ \hline
\endfoot
\hline
\endlastfoot
Environment &  &  &  \\
\hline
Minimum air temperature [°C] & Numerical & Spatiotemporal & BioSIM \\
Mean air temperature [°C] & Numerical & Spatiotemporal & BioSIM \\
Maximum air temperature [°C] & Numerical & Spatiotemporal & BioSIM \\
Total precipitation [mm] & Numerical & Spatiotemporal & BioSIM \\
Dew point temperature [°C] & Numerical & Spatiotemporal & BioSIM \\
Relative humidity [\%] & Numerical & Spatiotemporal & BioSIM \\
Solar radiation [watt/m2] & Numerical & Spatiotemporal & BioSIM \\
Atmospheric pressure [hPa] & Numerical & Spatiotemporal & BioSIM \\
Wind speed at 2 m [km/h] & Numerical & Spatiotemporal & BioSIM \\
Degree days [°C] & Numerical & Spatiotemporal & \\
Elevation [m] & Numerical & Spatial & BioSIM \\
Surface area [m2] & Numerical & Spatial & AA \\
Shoreline [m] & Numerical & Spatial & AA \\
\hline
Socioeconomics &  &  &  \\
\hline
Human population size in surrounding area [people] & Numerical & Spatial & StatCan (year 2021) \\
Mean income in surrounding area [CA\$] & Numerical & Spatial & StatCan (year 2021) \\
Median income in surrounding area [CA\$] & Numerical & Spatial & StatCan (year 2021) \\
Distance to next urban area [m] & Numerical & Spatial & AA, StatCan (year 2021) \\
Distance to road [m] & Numerical & Spatial & AA, StatCan (year 2021)\\
Change in work hours due to Covid-19 [\%] & Numerical & Temporal (quarterly, from Q4 2019 to Q4 2021) & StatCan \\
Average hourly wages & Numerical & Temporal (monthly), until January 2022 & StatCan \\
Consumer price index & Numerical & Temporal (monthly, until January 2022) & StatCan \\
Covid cases in the last seven days & Numerical & Spatiotemporal (Province) & \cite{Berry2021} \\
\hline
Fisheries management and events &  &  &  \\
\hline
Bag limitations & Boolean & Spatial & \cite{Ontario2019}\\
Fish size limitations & Boolean & Spatial & \cite{Ontario2019}\\
Catch-and-release regulation & Boolean & Spatial & \cite{Ontario2019}\\
Lake closure & Boolean & Spatiotemporal & \cite{Ontario2019}\\
Weekend day or weekday & Boolean & Temporal & - \\
Public holiday (+ connected weekend) & Boolean & Spatiotemporal & \url{https://www.statutoryholidays.com/}\\
Stocking event in the year & Boolean & Spatiotemporal &  Ministry Ontario (April 1, 2021) \\ %https://geohub.lio.gov.on.ca/datasets/793ddc06b9434694b99605edca233f89_0/explore?location=-0.000000%2C0.000000%2C5.83, https://www.gofishbc.com/stocked-fish/?rel_year=2022&reportType=regional 
Weeks since the last stocking event [weeks] & Numerical & Spatiotemporal & \\
\end{longtable}

\spacingset{1}
\setlength{\tabcolsep}{1pt}
\begin{table}[htbp]
    \begin{threeparttable}
    \caption{\added{Additional performance scores (Precision, Recall and F1-scores) of the best performing ML methods for predicting boat absence and presence.
    See SI Methods for further information on the performance scores. 
    Comparison of average performance of models using only the feature ``Website visits'', and models using features from Table \ref{tab:features} including, and excluding angler-reported data and ``Website visits''. 
    Spatio-temporal predictions were made at same lakes as used in model training (``known lakes", random training-test splitting) and at lakes that were unkown for the models (``unknown lakes", independent lakes splitting). 
    The mean was taken over the five models trained over different training-test data splits, respectively.}}
    \label{tab:add_pred_performance}
    \centering
    \scriptsize
    \begin{tabular}{|l|l|cc|cc|cc|cc|cc|cc|}
        \hline
        & & \multicolumn{4}{c|}{Only ``Website Visits''} & \multicolumn{4}{c|}{Including angler-reported} & \multicolumn{4}{c|}{Excluding angler-reported} \\
        & & \multicolumn{4}{c|}{} & \multicolumn{4}{c|}{data and ``Website Visits''} & \multicolumn{4}{c|}{data and ``Website Visits''} \\
        \cmidrule(lr){3-6} \cmidrule(lr){7-10} \cmidrule(lr){11-14}
        & & \multicolumn{2}{c|}{Training set} & \multicolumn{2}{c|}{Test set} & \multicolumn{2}{c|}{Training set} & \multicolumn{2}{c|}{Test set} & \multicolumn{2}{c|}{Training set} & \multicolumn{2}{c|}{Test set} \\
        & & Absence & Presence & Absence & Presence & Absence & Presence & Absence & Presence & Absence & Presence & Absence & Presence \\ \hline
        \textbf{Precision} & & & & & & & & & & & & & \\
        known & RF & 0.752 & 0.805 & 0.752 & 0.804 & 1.000 & 1.000 & 0.809 & 0.820 & 1.000 & 1.000 & 0.818 & 0.823 \\
        lakes & GBRT & 0.752 & 0.805 & 0.752 & 0.804 & 0.869 & 0.882 & 0.811 & 0.822 & 0.876 & 0.876 & 0.818 & 0.819 \\
         & SVM & 0.752 & 0.804 & 0.753 & 0.805 & 0.827 & 0.855 & 0.785 & 0.813 & 0.831 & 0.840 & 0.795 & 0.804 \\
        unknown & RF & 0.752 & 0.805 & 0.749 & 0.803 & 1.000 & 1.000 & 0.772 & 0.786 & 1.000 & 1.000 & 0.773 & 0.767 \\
        lakes & GBRT & 0.752 & 0.805 & 0.749 & 0.803 & 0.875 & 0.882 & 0.773 & 0.784 & 0.881 & 0.876 & 0.760 & 0.776 \\
         & LogReg & 0.707 & 0.855 & 0.712 & 0.860 & 0.769 & 0.823 & 0.742 & 0.787 & 0.773 & 0.796 & 0.746 & 0.772 \\ \hline
        \textbf{Recall} & & & & & & & & & & & & & \\
        known & RF & 0.810 & 0.746 & 0.809 & 0.745 & 1.000 & 1.000 & 0.811 & 0.817 & 1.000 & 1.000 & 0.813 & 0.828 \\
        lakes & GBRT & 0.810 & 0.746 & 0.809 & 0.745 & 0.878 & 0.874 & 0.813 & 0.819 & 0.869 & 0.883 & 0.807 & 0.828 \\
         & SVM & 0.808 & 0.746 & 0.809 & 0.746 & 0.853 & 0.829 & 0.807 & 0.789 & 0.832 & 0.838 & 0.793 & 0.804 \\
        unknwon & RF & 0.809 & 0.745 & 0.806 & 0.746 & 1.000 & 1.000 & 0.771 & 0.786 & 1.000 & 1.000 & 0.747 & 0.793 \\
        lakes & GBRT & 0.809 & 0.745 & 0.806 & 0.746 & 0.876 & 0.881 & 0.771 & 0.787 & 0.868 & 0.888 & 0.766 & 0.772 \\
         & LogReg & 0.883 & 0.651 & 0.889 & 0.656 & 0.828 & 0.762 & 0.795 & 0.734 & 0.790 & 0.779 & 0.770 & 0.751 \\ \hline
        \textbf{F1 Score} & & & & & & & & & & & & & \\
        known & RF & 0.780 & 0.774 & 0.778 & 0.773 & 1.000 & 1.000 & 0.809 & 0.817 & 1.000 & 1.000 & 0.814 & 0.824 \\
        lakes & GBRT & 0.780 & 0.774 & 0.778 & 0.773 & 0.874 & 0.878 & 0.811 & 0.819 & 0.872 & 0.879 & 0.811 & 0.822 \\
         & SVM & 0.779 & 0.774 & 0.779 & 0.773 & 0.839 & 0.842 & 0.794 & 0.799 & 0.831 & 0.839 & 0.792 & 0.803 \\
        unknown & RF & 0.780 & 0.774 & 0.776 & 0.773 & 1.000 & 1.000 & 0.771 & 0.785 & 1.000 & 1.000 & 0.759 & 0.779 \\
        lakes & GBRT & 0.780 & 0.774 & 0.776 & 0.773 & 0.875 & 0.881 & 0.772 & 0.785 & 0.874 & 0.882 & 0.762 & 0.773 \\
         & LogReg & 0.785 & 0.739 & 0.788 & 0.741 & 0.797 & 0.792 & 0.766 & 0.757 & 0.781 & 0.787 & 0.756 & 0.760 \\ \hline
    \end{tabular}
    \begin{tablenotes}
        \item RF- Random Forest, GBRT - Gradient-Boosted Regression Trees, SVM - Support Vector Machine, LogReg - Logistic Regression
    \end{tablenotes}
   \end{threeparttable}
\end{table}
\spacingset{2}
 
\end{document}